\documentclass[english,aps,pra,nofootinbib,twocolumn,superscriptaddress]{revtex4-1}
\usepackage{amsmath,amssymb}
\usepackage{bm,bbm}
\usepackage{color}
\usepackage{hyperref}
\usepackage{graphicx}
\usepackage{fdsymbol}
\usepackage{setspace}
\usepackage[T1]{fontenc}
\usepackage[latin9]{inputenc}
\newcommand{\1}{{\rm 1\hspace{-0.9mm}l}}

\begin{document}
\title{Five open problems in theory of quantum information}

\author{Pawe\l{} Horodecki}

\affiliation{International Centre for Theory of Quantum Technologies (ICTQT), University of Gda\'nsk, 80-308 Gda\'nsk, Poland}

\affiliation{Faculty of Applied Physics and Mathematics, Technical University
of Gda\'{n}sk, 80-952 Gda\'{n}sk, Poland}

\affiliation{National Quantum Information Centre (KCIK), University of Gda\'nsk, 81-824 Sopot, Poland}

\author{\L{}ukasz Rudnicki}

\email{lukasz.rudnicki@ug.edu.pl}

\affiliation{International Centre for Theory of Quantum Technologies (ICTQT), University of Gda\'nsk, 80-308 Gda\'nsk, Poland}

\affiliation{Center for Theoretical Physics, Polish Academy of Sciences, Aleja
Lotnik{\'o}w 32/46, 02-668 Warsaw, Poland}

\affiliation{National Quantum Information Centre (KCIK), University of Gda\'nsk, 81-824 Sopot, Poland}

\author{Karol \.{Z}yczkowski}

\affiliation{Faculty of Physics, Astronomy and Applied Computer Science, Jagiellonian University, 
30-348 Krak{\'o}w, Poland}

\affiliation{Center for Theoretical Physics, Polish Academy of Sciences, Aleja
Lotnik{\'o}w 32/46, 02-668 Warsaw, Poland}

\affiliation{National Quantum Information Centre (KCIK), University of Gda\'nsk, 81-824 Sopot, Poland}

 \date{December 21, 2020}

\begin{abstract}

We identify five selected open problems in the theory of quantum information, which
are rather simple to formulate,  were well-studied in the literature, but are technically not easy.
As these problems enjoy diverse mathematical connections,
they offer a huge breakthrough potential.
The first four concern existence of certain objects relevant for 
quantum information, namely 
a family of symmetric informationally complete generalized measurements
in an infinite sequence of dimensions,  mutually unbiased bases in dimension six, 
absolutely maximally entangled states for four subsystems with six levels each and bound entangled states with negative partial transpose. 
The fifth problem requires checking whether a certain state of  a two-ququart system is $2$-copy distillable.
 An award for solving each of them is announced.

\end{abstract}

\maketitle

{\sl Dedicated to memory 

of Roman Stanis\l aw Ingarden (1920-2011)

 on his centennial birthday 
}

\section{Introduction}
Roman Stanis{\l}aw Ingarden, one of the founding fathers of the field, wrote in 1975:
`The aim of the present paper was only to give a general formulation
of the quantum information theory of the Shannon type.
 The theory requires further investigations and mathematical development'.
  At the time this paper  {\sl Quantum information theory}  \cite{In76} 
  was written, exactly 45 years ago, 
  it was difficult to predict that such a piece of research in mathematical physics could
  inspire a vast new field of science and trigger a remarkable progress in experimental physics
  and yield numerous applications. 
 
Indeed, the field of quantum information 
 (see \cite{BDV00,NC10}) with its cornerstones of  pioneering discoveries of quantum money \cite{Wiesner83}, quantum cryptography \cite{Bennett-Brassard84,Ekert91}, quantum dense coding \cite{Bennett-Wiesner92}, quantum teleportation \cite{Bennett-Tele93}, quantum information compression \cite{Schumacher95} and quantum computing \cite{Deutsch-Jozsa92,Shor94,Grover97} has visibly matured 
in recent days, therefore, more and more often we hear and read about quantum technologies.
The latter aim at turning famous theoretical concepts such as
quantum cryptography into fully operational devices.
  These application, based on `standard' technologies developed so far, 
  possess essential functionalities solely operating on quantum principles.  

Along the course of the, so called, Second Quantum Revolution, experimental efforts are mostly directed towards solutions to practical problems, such as mitigation of the noise and decoherence effects, or scalability. Therefore, a discussion of perspectives within experimental quantum information could certainly focus on new techniques allowing for a better protection and control of quantum systems. 

On the theory side we observe a similar tendency. Current research focus is on optimization of theoretical protocols and experimental schemes, as well as discussion of practical limitations of the techniques developed. An example of a very recent, beautiful result \cite{crosstalk} from the field of quantum metrology can serve us the purpose of illustrating the above trend. While it is known that, so called, superresolution techniques \cite{superresolution} allow one to increase the precision beyond that of typical diffraction-limited direct imaging, robustness of this method is not fully understood. In  \cite{crosstalk}, a scheme based on intensity measurements involving spatial mode decomposition has been scrutinized against experimental noise stemming from the crosstalk between the modes used. Deterioration of the quantum superresolution benefits has been found. Based on the above example it is easy to imagine a perspective article devoted to theoretical quantum information, pointing areas and problems, within all the pillars of quantum technologies\footnote{For instance, within European Quantum Flagship initiative, quantum technologies are split into the following five pillars: quantum communication, quantum simulations, quantum sensing and metrology, quantum computing, quantum basic science.}, which require further attention.

In our manuscript, however, we take a perspective visibly different from that described above. Being aware of the currently relevant, particular challenges of theoretical quantum information, we ask ourselves whether there is still room for ground breaking, though not completely unexpected, developments. To let this question have an affirmative answer, we identify open problems with such a breakthrough  potential. We require the problems to be:
\begin{itemize}  
\item  well-studied and extensively covered by the topical literature,
so that a convincing evidence of their importance exists;
\item technically hard, so that they require methodology beyond the toolbox available at the moment;
\item universal and with a rich mathematical underpinning, so that they are not associated with narrowly defined platforms or protocols.  
\end{itemize}  
The first criterion assures the recognition, the proposed problems have gained. Being well-studied implies that the problems have a long history deeply immersed in the field of (theoretical) quantum information, therefore, a future solution shall expand the base of the field, rather than one of its distant branches. The second criterion, beyond offering an explanation why the problems still remain open, pertains to the future impact of the solutions. Presumably advanced techniques necessary to tackle the problems, perhaps not yet recognized or even established, will likely make an impact beyond their initial niche. Finally, since we observe the tendency towards specialization and narrowing of the research conducted, trends which are essential at a stage where initially broad concepts are being turned into concrete devices, we look for breakthrough theoretical discoveries beyond this \textit{modus operandi}. Likely, only unexpected solutions to problems which are not associated with a particular setup can influence the whole field of quantum information.

In our tight selection of the open questions to be offered as a future inspiration and guideline for theoretical research, we restrict our attention to five concrete problems. 
Why five? 
 As it turns out, the number six,  if used to set the dimension of the Hilbert space, 
 is still insufficiently well understood in context of quantum information.
While the above
 justification gives as good a reason as any other reason, the first three problems described below are in fact associated with symmetric configurations in discrete Hilbert spaces, 
 and two of them are to some extent concerned with this special dimension. 

\section{Discrete structures in the Hilbert space}

The space of pure quantum states of a fixed dimension\footnote{
While presenting the problems we used notation common in each
subfield. Thus the reader is advised that the notation is not entirely
 consistent throughout the entire paper. Consequently,
in different problems the symbols $n, d, k$ and $N$ have different
meaning.} $N$
is isotropic -- no quantum
state is `more equal than others'. However, this property does not exclude existence of complex and at the same time well-organized structures inside the Hilbert space, e.g. particular {\sl constellations} of quantum states with prescribed properties. It is easy to imagine, that each structure of such kind nurtures a potential for quantum information protocols such as those used in error correction, or particular experimental tasks, such as quantum tomography. Readers familiar with the background of theoretical quantum information will likely recognize that {\sl mutually unbiased bases} (MUBs) and {\sl symmetric informationally complete positive operator valued measures} (SIC POVMs) provide natural examples of such structures. Intriguingly, in both cases there is an important missing piece of the puzzle which we now turn into an open problem. Additionally, we extend our discussion to cover the third constellation, perhaps better known in classical considerations (therefore, below we introduce it in more detail), namely the {\sl Latin squares} (LS). 

Following the common word of wisdom saying that the proof of the pudding is in the eating we shall cut here the general discussion and  immediately pose the three problems associated with the three constellations/structures mentioned above.

\subsection{Existence of SIC POVMs}
{\bf Problem 1:} {\sl  Construct SIC POVMs in an infinite sequence of dimensions,
   $N_1, N_2, N_3, \dots$}

\bigskip

\begin{figure}[h]
	\centering
	\includegraphics[width=0.83\linewidth]{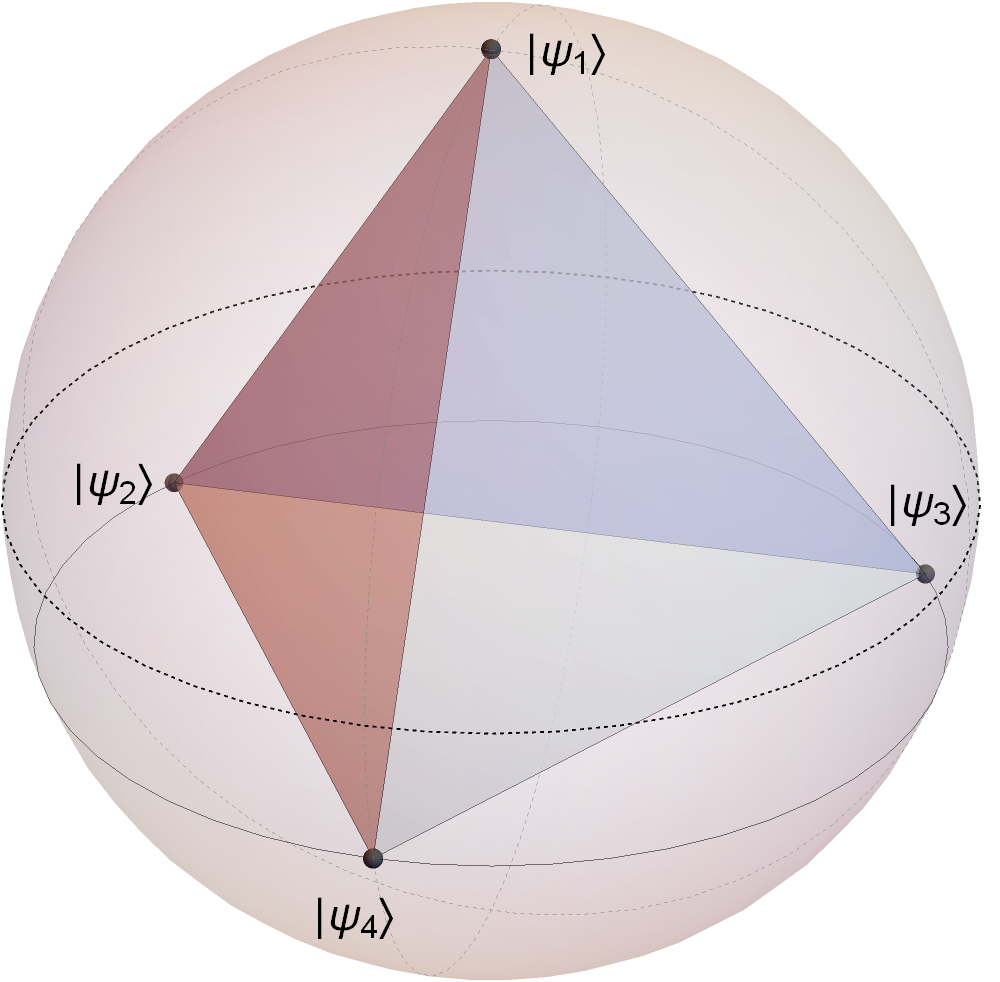}
	\caption{Four pure states $|\psi_i\rangle$ 
	  span a regular tetrahedron inscribed in the Bloch sphere
	  and lead to a single-qubit  symmetric informationally complete measurement (SIC POVM)
	    for $N=2$.
	   Can you find $N^2$ pure states of size $N$ such that 
	   the corresponding projectors form a simplex
	   inscribed into the set of quantum states of a given order $N$? }
 \label{fig:SIC2}
\end{figure}

{\bf  Setup.} 
A symmetric informationally complete positive  operator valued measure  \cite{Za99,RBSC04} 
associated with an $N$-dimen\-sional complex Hilbert space ${\cal H}_N$
is given by a set of $N^{2}$  
vectors $|\psi_j\rangle \in {\cal H}_N$
satisfying  the following overlap relations, 
\begin{equation}
|\langle \psi_j|\psi_k\rangle |^2=\frac{N \delta_{jk} + 1}{N + 1},
\ \ \  \ j,k=1,\dots, N^2.
\end{equation}
This set defines a generalized quantum measurement
capable to extract complete information concerning any density matrix of order $N$
described by $N^2-1$ real parameters. 
Moreover, such a constellation of $N^2$ projectors onto pure states 
forms a simplex inscribed in the entire $N^2-1$ dimensional set of 
density matrices of order $N$ -- see Fig. \ref{fig:SIC2} for an $N=2$ example.
For an  accessible guide to the SIC problem in low dimensions 
consult   \cite{An15}. 

\medskip
{\bf Motivation.}
From a mathematical point of view, we ask about the maximal 
set of {\sl complex equiangular lines} \cite{Stacey20} in a given dimension $N$.
From a physical perspective one looks for a scheme of an optimal
quantum measurement of an arbitrary size $N$,
distinguished by the fact that the number of projector operators
is minimal possible required 
to gather complete information concerning the analyzed state.
Solving the SIC existence problem for any dimension
will significantly contribute to our understanding of the
 set-of-quantum-states' geometry  \cite{BZ17}.

According to the 1999 dated conjecture by Zauner \cite{Za99},
 for any dimension $N$ there exists a fiducial vector,
such that all remaining $N^2-1$ elements of the desired SIC
can be obtained by acting on it with unitary matrices
representing elements of the  Weyl--Heisenberg group.

Numerical solutions obtained in 2004 for all dimensions up to  $N =45$  (Renes et al. \cite{RBSC04}) were
extended  in 2010   by Scott and Grassl \cite{SG10} to  $N \le 67$.
Further results from 2017 included dimensions
$ N\le 121$, Scott  \cite{Sc17} and  $ N\le 151$,  Fuchs et al.  \cite{FHS17}.
In 2020 numerical solutions were known for
$ N \le 193$ and also for
$ N = 204, 224, 255, 288, 528, 725, 1155, 2208$ (Grassl  \cite{Gra20}).
Analitical solutions are known  
for $N\le 53$ \cite{RBSC04,SG10,Sc17,ACFW18},
and several other dimensions, including
 $N=57, 61-63, 65, 67, 73, 74, 76, 78-80, 84, 86, 91, 93, 95$,
$ 97-99, 103, 109, 111, 120, 122, 124, 127, 129, 133, 134, 139$, 
$143, 146, 147, 151, 155, 157, 163, 168, 169, 172,181-183$, 
$193, 195, 199, 201, 228, 259, 292, 323, 327, 364, 399, 403, 489$,
$ 844, 1299$ -- see \cite{GS17, AB19, Gra20}.

However, in spite of a considerable  research effort 
   \cite{Zhu10, JW15, ABDF17,AD19,IngemarNEW},
the general conjecture of Zauner remains unproven.
Finding a family of SICs in any infinite sequence of dimensions
could become a decisive step in this direction.
Furthermore, let us emphasize inspiring connections
to some major open questions in algebraic number theory, 
including a key part of  the  12-th
 problem of Hilbert \cite{AHAZ13, AFMCY17,Ko18,Ingemar2020}.

\subsection{MUBs in dimension six}

{\bf Problem 2:} {\sl Construct a set of {\sl at least} $4$  mutually unbiased bases
      in dimension six or prove that  there are no $7$ MUBs in ${\cal H}_6$.}                               

\bigskip

\begin{figure}[h]
	\centering
	\includegraphics[width=0.92\linewidth]{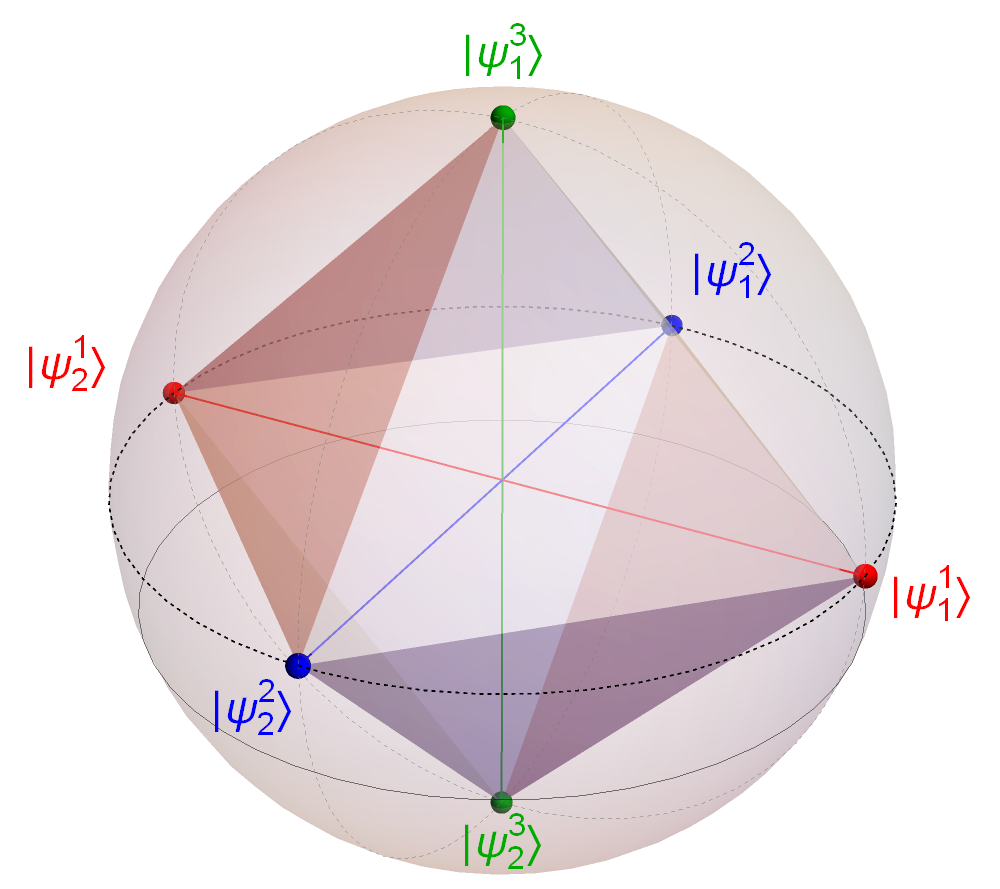}
	\caption{Three eigenbases of Pauli matrices $\sigma_x,\sigma_y,\sigma_z$
	 span an octahedron inscribed in the Bloch sphere and form a set of $3$ 
	 mutually unbiased bases for $N=2$. There exist $4,5,$ and $6$ MUBs in dimensions
	$N=3,4$ and $5$, respectively. How many MUBs there exist for $N=6$?}
 \label{fig:MUB2}
\end{figure}

{\bf  Setup.} 
Consider a set of $K$ bases  $\{ |\psi_{i}^{m}\rangle \}$
 ($1 \leq m \leq K$, $1 \leq i \leq N$) in $N$-dimensional complex
 Hilbert space ${\cal H}_N$,
 so that all  vectors in each basis are orthogonal,
$\langle  \psi_{i}^{m} |  \psi_{j}^{m} \rangle = \delta_{ij}$.
 These bases are called mutually unbiased
if any two bases are unbiased, which means
 \begin{equation}
\forall_{i,j} \qquad |\langle  \psi_{i}^{m} |  \psi_{j}^{n} \rangle|^2 = \frac{1}{N}, \qquad m\neq n.
\end{equation}
It is relatively easy to show that  there exist  no more than $N+1$  MUBs in ${\cal H}_N$.
Moreover, for any $N\ge 2$, there exist at least three MUBs (see Fig. \ref{fig:MUB2} for an $N=2$ example).
If the dimension $N$ is a prime number or a power of a prime, $N=p^k$,
there exists a complete set of $N+1$ MUBs  \cite{Iv81,Wo89}.
This implies that for a composite dimension represented 
by a product of powers of primes, $N=p_1^{k_1}\dots p_m^{k_m}$,
with $p_1^{k_1} \le p_2^{k_2} \le \dots \le p_m^{k_m}$,
there exist (at least) $p_1^{k_1}+1$ MUBs \cite{Za99,KR04}.
It is also known that if one finds $N$ MUBs in dimension $N$
the last $(N+1)$-th unbiased basis also exist \cite{We13},
so the maximal number of existing MUBs is either equal to $N+1$,
or it is less or equal to $N-1$. 

Several methods to construct MUBs are known \cite{BBRV02,KR04,Ar05} 
 and all solutions for dimensions $2-5$ are classified \cite{BWB10}.
If $N$ is a power of a prime, various properties of a complete set of $N+1$ MUBs
 are already understood \cite{La04,RBKS05,Co09, WPZ11,MPW16},
 but otherwise 
 the number of existing MUBs remains  unknown  \cite{WB05,DEBZ10, BC18}.
In particular, for $N=6$ a complete set would consist of seven MUBs,
but up till now only solutions containing three bases 
were found \cite{BBELTZ07, BH07, BS09,Gr09, JMMSW09, JMM10,RLE10, Go13,BFNE16, CY18}.
 It is however known that if a complete set of seven MUB exists, 
it cannot contain a triple of product bases \cite{MW12b,MW12c}.

Any unitary matrix  which relates two unbiased bases,
$U_{ij}=\langle  \psi_{i}^{m} |  \psi_{j}^{n} \rangle$,
belongs to the set of complex Hadamard matrices.
This set consists of unitary matrices of order $N$, such that
all its entries have the same squared modulus, $|U_{ij}|^2=1/N$.
Interestingly, the set of complex Hadamard matrices is again fully characterized
 \cite{Ha96,TZ06,Ba19} up to  $N= 5$.
 As several new complex Hadamard matrices of order $N=6$
 were discovered  a  decade ago  \cite{Sz10,Ka11},
 it was tempting to expect that they could lead to certain
 sets of four MUBs in this dimension \cite{Go13}. 
However, up till now the maximal number of 
MUBs for $N=6$ remains three,
even though for larger dimensions further connections 
between Hadamard matrices and MUBs were found \cite{MW12d}.
             
\medskip             
 {\bf Motivation.} On the one hand, finding a complete set of MUBs
  in dimension $6$ would yield an optimal scheme of orthogonal 
  quantum measurement in this dimension. 
 More importantly, deciding whether such a configuration exists 
 has significant implications for foundations of quantum theory,
 as up till now our understanding of basic properties
 of finite dimensional Hilbert spaces is not complete.
 On the other hand, a possible non-existence result is of a considerable mathematical interest,
 as it would provide further arguments that the number $6$ -- the smallest product of two different primes --
   is indeed very special and `less equal than others'.
   Let us emphasize here that there is no affine plane of order six
   and there are no orthogonal Latin squares of order six, which inspires the subsequent problem.
Research on the MUBs reveals further  intricate links between
foundations of quantum theory
  and  several  fields of mathematics,   including  Galois rings,  group theory,
 combinatorics, finite fields and projective geometry \cite{SPR04,  BE05, KP06,PRP06,BSTW07,
PDB09, PPGB10, KMW18}.

\subsection{Quantum Orthogonal Latin Squares}

{\bf Problem 3:}
{\sl Determine whether there exist two quantum orthogonal Latin squares 
\cite{GRMZ18,MV19} of order six.  In other words,
find a solution of the problem of \; $36$ `entangled officers' of Euler
or demonstrate that it does not exist. }

\bigskip

{\bf Setup.} 
    A Latin square of order $N$ is filled with $N$ copies of $N$ symbols
    arranged in a square in such a way that no row nor column of the square
   contains the same symbol twice. 
   The name refers to papers of Leonhard  Euler \cite{Eu1782},
    who used Latin characters as symbols to be arranged. 
   To enjoy a simple example use the Pauli matrix and write down 
   $2\sigma_x+{\mathbbm 1}_2$. It is likely 
    that the Euler's approach to the problem was rather different...
   
   Two orthogonal Latin squares (also called Graeco-Latin squares)
   of order $N$ consist of $N^2$ cells arranged in a square
   with a pair of ordered symbols in each cell,
   for instance one Greek character and one Latin. 
   Every row and every column of the square
   contains each possible pair of symbols exactly once,   
    and  no two cells contain the same ordered pair -- see Fig.  \ref{fig:MOLS4}.
    A set of $k$ Latin squares which are pairwise orthogonal
    are called {\sl mutually orthogonal Latin squares} (MOLS). 
    It is easy to show that for a given  $N$  there exist no more than 
    $N-1$  MOLSs. 
    Similarly with the case of the MUBs, this bound is saturated if 
    $N$ is a prime or a power of a prime  \cite{DK91}.  
   
Historically,   Euler analyzed the problem of  $36$ officers  
   from six regiments, each containing $6$ officers of $6$ different ranks. 
    They should be arranged before a parade into a 
    $6 \times 6$ square such that each row and each column
    holds only one officer from each regiment and only one officer from each rank.
   Euler wrote in 1782 that this problem has no solution  \cite{Eu1782}
    without providing a formal 
   proof, established  only in 1901 by Gaston Tarry \cite{Ta01}.
   This result implies that there is no pair of orthogonal Latin squares of size $6$, 
   so that the upper bound for the number of MOLS, in this case  $N-1=5$, is not saturated.
   For any  $N\ge 7$ there exist at least two MOLS,
   in particular also for $N=2 \times 5=10$  -- consult a novel by Georges Perec \cite{Pe78}.
    In general,  the problem of finding the maximal number of MOLS for an arbitrary value of $N$ remains open 
    \cite{CD01}.

\begin{figure}[h]
	\centering
\begin{equation} { 
\huge
 \begin{array}{|cccc|} \hline 
{ A\spadesuit}  & K\clubsuit&  {\color{red} Q\diamondsuit }&  {\color{red} J\varheartsuit}  \\ 
{ {\color{red} K\varheartsuit}} & {\color{red} A\diamondsuit }& { J\clubsuit}& Q\spadesuit\\  
{Q\clubsuit}    & J\spadesuit  &{{\color{red}A\varheartsuit}} & {\color{red} K\diamondsuit} \\  
{\color{red} J\diamondsuit} & {\color{red} Q\varheartsuit} & K\spadesuit & A\clubsuit \\ \hline
\end{array} } 
\nonumber
 \end{equation}
	\caption{An example of $N = 4$ Greaco-Latin square prepared 
	for bridge players \cite{BP}.
Due to works of Euler and  Tarry we know that for $N=6$
a similar design of 36 cards of 6 different suits and 6 different ranks
(or 36 officers of different ranks and arms) does not exist.
Is there a solution of the  $N=6$ problem if we play bridge with quantum cards, like
 $ (|K\clubsuit \rangle + | {\color{red} Q\vardiamondsuit}\rangle)/\sqrt{2}$,
 or allow the officers of Euler to be entangled?}
 \label{fig:MOLS4}
\end{figure}

As a rule of thumb, for any interesting classical notion
one can find a quantum analogue.  
A {\sl quantum Latin square} is an $N \times N$ table of  $N^2$ vectors from 
$N$ dimensional Hilbert space ${\cal H}_N$ 
arranged in such a way that every row and every column of the table 
forms an orthonormal basis in the space \cite{MV16}.  
{\sl Orthogonal quantum Latin squares} (OQLS) are defined \cite{GRMZ18}
as  a collection  of  $N^2$  normalized vectors from a composite  space  
${\cal H}_N \otimes {\cal H}_N$,
 which are  mutually orthogonal so they form an orthonormal basis.
They are arranged in an $N \times N$ table such that for every row (column)
 the symmetric superposition of all states in each row (column) 
 is a maximally entangled state, for instance 
\begin{equation}\label{maxent}
 |\psi_+\rangle= \frac{1}{\sqrt{N}} \sum_{j=1}^N |j\rangle \otimes |j\rangle.
\end{equation}
 Any  Graeco-Latin square leads to  such a design,
 since it suffices to treat the pair of classical objects $(\alpha,  B)$ 
 as a product state, $|\alpha \rangle \otimes |B \rangle$.

It is known that the generalized Bell state (\ref{maxent})
is maximally entangled among all states of a 
bipartite $N \times N$ system.
For other systems, a natural question arises
 \cite{GBP98,HS00,KNM02, Sc04,BSSB05,BP+07,FFPP08,MGP10,FFMPP10b,AC13}:
What are the most entangled states for quantum systems consisting of $N$ systems
with $d$ levels each ? 
The answer depends on the entanglement measure used \cite{BZ17},
but already for a four-qutrit system there exist a state, 
which displays maximal entanglement with respect 
to all three possible splittings of the entire system into two pairs of qutrits.
Such a state is
called {\sl absolutely maximally entangled} (AME) \cite{HCLRL12}.

This notion can also be generalized for larger systems.
An $n$-partite pure state is  called AME state  
if it is maximally entangled with respect all possible bipartitions \cite{HCLRL12},
so that all its reductions consisting of  $k$  subsystems, 
with arbitrary  $k \leq   \lfloor {n}/{2} \rfloor$,  are maximally mixed. 
A density matrix  on 
a given $N$-dimensional Hilbert space is {\it maximally mixed},
if it is proportional to the identity operator 
on this space, $\rho_{*}={\1}_N/{N}$.
It is known that there are no AME states of 4 qubits \cite{HS00},
and equivalently,  a pair of  QOLS does not exist for $N=2$.
Scott demonstrated  \cite{Sc04} that in the case 
systems of $m$ qubits the AME states so exist for $m=3,4,5,6$
and also showed that they do not exist for $m\ge 8$.
The last remaining issue of  $m=7$ qubits was later solved in 
by Huber et al. \cite{HGS17},
who proved that such AME states do not exist.
The list of currently known AME states is available on line \cite{HW_table}.

As there are no two orthogonal Latin squares (OLS)
of order six, the famous classical problem of {\sl 36 officers of Euler}
 has no solution \cite{CGSS05}. An analogous quantum problem,
which involves $36$ {\sl entangled officers},  remains open. 
\medskip

{\bf Motivation.}
This problem can be reformulated in several  other settings.
To present them we need to review some further notions.
A unitary matrix $U$ of size $N^2$ is called $2$-unitary \cite {GALR15}
if both the partially transposed matrix $U^{\Gamma}$
and the reshuffled matrix $U^R$  remain unitary -- for the definitions 
of these reorderings of the entries of a matrix
see (\ref{NPT}) and consult \cite{BZ17}.

Any matrix of a square size can be represented as a tensor 
$T_{ijkl}$ with four indices.
   Such a tensor can be reshaped into a matrix $X_{\mu\nu}$
  using composed indices in three different ways:
   a) $\mu=\mu(ij)$, 
   b) $\mu=\mu(ik)$  and c) $\mu=\mu(il)$,
     while the second index  $\nu$ is obtained in each case  from the
      remaining two indices.
     A tensor $T$ is called {\sl perfect} \cite{PYHP15} 
    if for any of these three ways of reshaping it, the outcome matrix becomes unitary. 
    Hence any  flattening of a perfect tensor forms a $2$-unitary matrix.
\smallskip

Establishing a negative result concerning existence of
two  quantum orthogonal Latin squares of order six is equivalent to
proving that 
\begin{itemize}

\item[i)] there is no AME state
       of four subsystems with six levels each \cite{HCLRL12,HESG18}
   thus the corresponding quantum error correction code  \cite{GBR04},
written  $((4,1,3))_6$, does not exist;
    \item[ii)]  there is no 2-unitary matrix $U \in \mathcal{U}(36)$;
       \item[iii)] there is no {\sl perfect tensor}
        with four indices, each running from $1$ to $6$. 
\end{itemize}

Furthermore, a negative result would directly imply the famous
    Euler conjecture that there are no two orthogonal Latin squares of order $6$.
    On the other hand, a positive result
    could become an important step towards development of {\sl quantum combinatorics}:
     a search for particular constellations of discrete quantum objects, 
     with special properties of symmetry and balance,
       hidden in the continuous Hilbert space. 
      As the standard combinatorics deals with discrete objects 
      and is related to the group of permutations,
      its quantum analogue concerns the continuous space of quantum states
      and relies on the continuous unitary group.
      
      Since  the problems number 2 and number 3 refer to the  same dimension
       (equal to their product), $N=6$,
        it is natural to speculate that they might be somehow related.
        It seems, however, that a  connection between
        problems of finding the maximal number of MOLSs
        and MUBs for a given dimension is not a direct one \cite{PDB09,PPGB10}.
        On the other hand several links between both problems were established: 
        Wocjan and Beth used (classical) MOLSs to construct a set of $6$ MUBs in dimension 
         $N=676$, which beats the prime power construction
          applied to the factorization $N=2^2 13^2$ yielding
            only $4+1=5$ MUBs \cite{WB05}.
            Furthermore,  Musto  used quantum Latin squares
            to construct in square dimensions mutually unbiased bases 
            consisting of maximally entangled states  \cite{Mu16}.

\subsection{Further perspectives I}

The three problems discussed above concern finite dimensional
Hilbert spaces. However, the notion of MUBs (or mutually unbiased measurements -- MUMs) is also present in experimentally relevant setups involving continuous \cite{WeigertCV} or coarse-grained \cite{MUMPRL} systems. While in the continuous case we maximally have $3$ MUBs \cite{WeigertCV}, in the coarse-grained scenario \cite{ManyMUMs} the situation is much more elaborate. Interestingly, the special dimension $N=6$ is not at all distinguished in the coarse-grained setting, since the systems of even dimension behave like the continuous ones (no more than 3 MUMs) --- only odd dimensions nurture potential for more \cite{ManyMUMs}. Whether this fact is connected to the conundrum of $N=6$ for discrete systems remains and open question at the moment.

Real Hadamard gates play a key role in numerous schemes of
 quantum information processing. More general, complex Hadamard matrices are instrumental in  {\bf Problem~2} concerning MUBs,
  but they become also linked  \cite{ABFG19}
  to {\bf Problem~1} on SICs. 
  These matrices do exist in any dimension -- as for any dimension $N$ we can write down 
  the Fourier matrix $F_N$ -- 
in contrast to real Hadamard matrices  \cite{DEBZ10}. 
They were constructed  by Sylvester  more than 150 years ago \cite{Syl67}, 
 but it was Hadamard who first showed \cite{Ha93} 
 that such matrices do not exist unless $N=2$ or $N=4k$.
  The celebrated conjecture due to Paley \cite{Pa33} 
 that they do exist for all dimensions not excluded by Hadamard is now
  confirmed \cite{KTR05} up to $N=664$. The point is that finding a solution for $N=4k$ gives us no clues,  whether a real Hadamard matrix exists for $N'=4k+4$. We encounter a
   similar situation 
 in {\bf Problem~1}, as finding SIC for a given $N$ sheds no light into the existence problem in dimension $N+1$. However, it is straightforward to
 construct  an infinite family  of Hadamard matrices in dimension 
 $N=2^m$ by tensor product, 
  while we are still in search for  a family of SICs in an infinite sequence of dimensions.
 Knowing a SIC configuration 
 for a certain dimension $N$ it might be easier to look for another one
 in dimension $N(N-2)$, but such a SIC dimension tower \cite{ABDF17,AD19}
 remains, up till now,  of a finite size only.

The first two problems deal with pure quantum states from a single space ${\cal H}_N$,
while  the last one requires to consider at least its two copies,
 ${\cal H}_N \otimes {\cal H}_N$.
Such a space with a tensor product structure
corresponds to a physical system in which two subsystems can be distinguished.
This construction allows one to introduce product states
and entangled states,  including the generalized Bell state (\ref{maxent}).

 {\bf Problem~3} concentrates on  the Hilbert space of dimension six,
but it is clear that more than a single copy of ${\cal H}_6$ has to be involved.
Looking for a solution of the generalized Euler problem
we are allowed to play with quantum cards,
 $\cos \varphi |A\spadesuit \rangle +  \sin \varphi | {\color{red}K\varheartsuit}\rangle$,
 and analyze configurations of $6^2=36$
   possibly entangled states in ${\cal H}_6 \otimes {\cal H}_6$.
 
 To create entanglement in a bipartite $N\times N$ system we need a global unitary gate,
 $U\in U(N^2)$, which couples both subsystems.
For any such a bipartite gate $U$ one defines an {\sl entangling power}
 \cite{ZZF00, Sc04,CGSS05} 
as the  average linear entropy of entanglement  \cite{BZ17}
created when U acts on a random product state
sampled according to the Haar measure on  both subspaces.
 
 A unitary matrix $U$ of size $N^2$, which saturates the absolute bound for the 
 entangling power,
 has to be $2$-unitary and therefore it
  allows one to construct the AME state for four subsystems
 with $N$-levels each -- see \cite{JMZL20}. 
In the latter formulation of the problem 
devoted to the search of distinguished AME pure states 
of four parties \cite{HCLRL12,HW_table},
one works with four subsystems with $6$ levels each,
 which are represented in the space  ${\cal H}_6^{\otimes 4}$.
   
 Interestingly, the same problem can also be mapped into 
 a question concerning properties of certain mixed states
 of the squared dimension.
 It was recently demonstrated \cite{YSWNG20}
 that existence of the desired AME state of four 
 subsystems with $6$ levels each is equivalent to the bipartite 
 separability of a certain mixed state $\rho$ living in 
  ${\cal H}_6^{\otimes 8}={\cal H}_6^{\otimes 4}    \otimes {\cal H}_6^{\otimes 4}$.
As the dimension of the problem becomes thus very high,
it is difficult to imagine that this approach 
could give us soon a constructive answer to the problem posed.

\medskip

The above remarks exemplify a common observation that various 
problems in one branch of mathematical physics 
can reveal unexpected links to questions in apparently distant fields of science.
As the last {\bf Problem~3} is shown to be closely related to
the various questions concerning entanglement of  mixed quantum states,
  we are now ready to proceed to the second part of the paper
   devoted exactly to these issues.
  
\section{Quantum entanglement \\ and its  distillability}

One of the most fundamental notions in the 
   theory of quantum information processing is that of entanglement. We say that a bipartite product state is called separable, 
 while all other pure states are entangled.
 A density matrix representing a mixed state is called  entangled
  if it cannot be represented as a convex combination of product states  \cite{Werner1989}.
  
  Entanglement proved itself to be a crucial resource relevant for quantum information processing.
 Therefore, one of the major problems in this field has been, since the early days of quantum information, to decide
 whether a given quantum state of a composite
system is separable or entangled \cite{HHHH09}.

Perhaps surprisingly, up till now,
 this general problem is solved only for $2 \otimes 2$ and $2 \otimes 3$ systems\footnote{The system is called $d \otimes d'$ 
 	iff one associates with it the Hilbert space with a tensor product structure, 
 	 ${\cal H}_d \otimes  {\cal H}_{d'}$.},
as in these cases  the single positive partial transpose  criterion provides a 
constructive answer \cite{HHH96}.
Already for a $3 \otimes 3$ system, neither 
a finite number of positive-maps-based separability criteria  \cite{Sk16} 
nor a techique using finite-size semi-definite programming \cite{Fa19} allows us to conclude whether a given quantum state is entangled or not.
 Moreover, the known procedure for deciding the separability 
 of a given bipartite quantum state in a finite number of steps  \cite{BHHA13} 
cannot be applied in practice due to its high complexity. 
 
The above problem, as well as the whole associated subfield concerned with certification (tests) or quantification (measures) of entanglement, gains a visibly less attention in recent years. This occurs likely because a lot has in fact been achieved, and it is relatively hard to identify research directions promissing an intellectual reward. Not to dream about giving a twist  to the whole subfield. Here we attempt to offer two such problems which, when solved, are capable of boosting the research on quantum entanglement \textit{per se}. 
 
\subsection{Bound entanglement}
{\bf Problem 4:} {\sl  Establish whether there
exist  bound entangled states with negative partial transpose.}                            
\medskip

{\bf   Setup.} 
To analyze quantum entanglement it is useful to introduce
the notion of {\sl partial transpose} of a density matrix.
Let $\rho$ denote a bipartite quantum state with matrix elements
written in a product basis, $\rho_{ij,lm}= \langle i j| \rho | l m  \rangle$.
Then its  {\it partial transposition},  $\rho^{\Gamma}$, reads
\begin{equation}\label{NPT}
\langle i j | \rho^{\Gamma} |  l m \rangle = \langle i m| \rho | l j  \rangle. 
\end{equation}
A  quantum state is said to have {\it positive partial transpose} (PPT) 
if all eigenvalues of its partial transpose are nonegative.
Otherwise, if some eigenvalues of  $\rho^{\Gamma}$ are strictly smaller than zero,
the state has  {\it negative partial transpose} (NPT).

  The concept of {\it entanglement distillation} refers to
protocols that allow to transform noisy entangled states to maximally 
entangled states in a well defined scenario. It was  originally discovered in 1996 when explicit
protocol was proposed for a class of mixed two-qubit states 
\cite{Bennett-Distillation96} and applied in its general form to quantum error correction 
\cite{Bennett-ErrorCorrection96} and cryptography \cite{Deutsch-EtAl-96}.  
Soon after it was shown that all two-qubit entangled states are distillable
\cite{HHH97}. However, in higher dimensional systems
some entangled states can be distilled to a singlet form \cite{HH99}, 
but  there exist also nondistillable entangled states,
which are called {\sl bound entangled}  \cite{HHH98,Ho19}.

If the dimension of a bipartite system is larger than six,
 there exist entangled states with PPT property \cite{Ho97}.
As all entangled states with PPT property are nondistillable,
 in these dimensions there exist bound entangled states \cite{HHH98}
 and the set of these states has a positive measure \cite{Zy99}.

 We analyze here {\bf Problem 4} with 
	equal dimensions of both subsystems, as the general case, $d\ne d'$,  can be 
	reduced to it. 
 A $d$-dimensional\footnote{We now on purpose denote the Hilbert-space dimension by $d$, while in the former section we used $N$, in order to follow the notation common in the subfield.} bipartite state $\rho$ defined on a composite
  Hilbert space   ${\cal H}_{d} \otimes  {\cal H}_{d}$  
is called  {\it distillable}, if it is {\it n-copy distillable} for some finite $n$.
 The property of $n$-copy distillability means that there exist two-dimensional (i.e. of rank two) projectors $P$ and $Q$ such that 
the matrix $(P \otimes Q) [\rho^{\Gamma}]^{\otimes n} (P \otimes Q)$ has 
a negative eigenvalue \cite{HHH98,Cl06,Ho19}.
It should be stressed that the projectors $ P$  and $Q$ act
 on the product $({\cal H}_d)^{\otimes n}$ of all  $n$ Hilbert spaces  associated with 
left and right subsystems of copies of the considered bipartite system respectively
--- see  \cite{DSSTT00,Pankowski2010}.

The question of  NPT bound entanglement  is closely related to a mathematical problem 
concerning 2-co-positive maps \cite{DSSTT00, Clarisse06PhD}. 
 A linear map 
$\Lambda: M_{d}({\cal C}) \rightarrow M_{d}({\cal C})$
acting on ${\cal H}_d$ is called {\it positive}
 iff it transforms any matrix with non-negative eigenvalues into a matrix with the same property. Furthermore, a linear map $\Lambda$ is called {\it $k$-positive} if and only if  the following extension  ${\1}_{k} \otimes   \Lambda : M_{k}({\cal C})   \otimes M_{d}({\cal C})  \rightarrow M_{k}({\cal C})   \otimes M_{d}({\cal C}) $ is positive, where  $
  {\1}_{k} $ stands for the identity map, which sends
   any  complex matrix from $M_{k}({\cal C})$ into itself.
   In particular $1$-positivity is equivalent to positivity. 
  The map is called {\it completely positive} iff it is $k$-positive for any $k$. 
  For a finite dimension $d$, to ensure complete positivity 
   it is enough to check only $k$-positivity for $k=d$. 
A map $\Lambda$ is called {\it $k$-co-positive} (resp. {\it completely co-positive} ) if and only if the composition
  $S= T  \circ \Lambda  $  is $k$-positive (resp. {\it completely positive}), where $T$ stands for transposition. 
  In particular, $1$-co-positive maps are called just {\it co-positive}.

\medskip

\begin{figure}[h]
	\centering
	\includegraphics[width=0.98\linewidth]{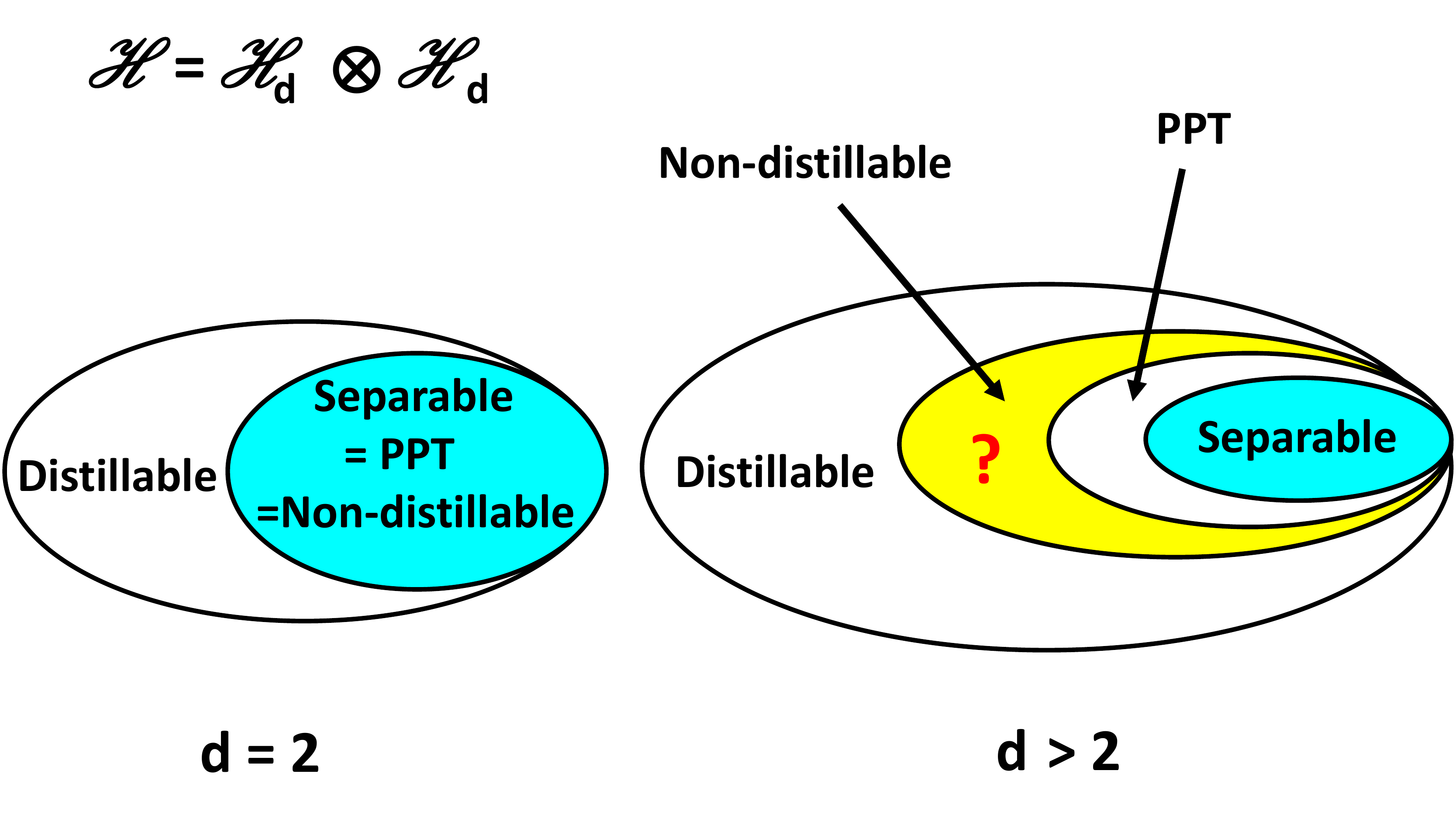}
	\caption{In the two-qubit problem, $d = 2$, 
	  the set of separable states coincides with the set of PPT states 
     and there are no bound entangled states since any entangled state is distillable.
	  For higher dimensions, $d > 2$, one asks whether the hypothetical region representing 
	       bound entangled states with  negative partial transpose,  
	         depicted in yellow,  is empty or not.}
	\label{fig:NPT-bound-entanglement}
\end{figure}

 {\bf Motivation.}                 
This is one of the long-standing open questions of quantum information theory \cite{DSSTT00, DCLB00}.
 It provides a sharp distinction between the two-qubit case, 
 in which all entangled states are distillable \cite{HHH97}, 
 and higher dimensional $d \otimes d$ problem, for which the question of existence of bound entangled states with negative partial 
 transpose is open -- see Fig. \ref{fig:NPT-bound-entanglement}.

Its positive solution would therefore have several consequences. If NPT bound entangled states exist then the set of non-distillable entangled states is neither closed under the tensor product  nor under mixing -- see \cite{Clarisse06PhD}. 
 The latter means that there would exist  two non-distillable entangled states  such that their mixture were distillable.  This  would imply  that one of the central measures of entanglement theory, namely distillable entanglement (which describes asymptotic amount of entanglement that can be distilled from many copies of a given state by local operations and classical communication \cite{HHHH09}) is neither additive nor convex \cite{Clarisse06PhD}.

  In this way, a possible affirmative solution to the present problem would lead to an
   extremal example of {\it superadditivity}. Namely, it has been proven that for any NPT state there exists PPT bound entangled state such that the product of the two is distillable \cite{Eggeling+2001}. Consequently, if the NPT state were bound entangled, then we would have the pair of two bound entangled states (both with their individual distillable entanglement measure equal to zero) such that their tensor product would be distillable (i.e. having the measure strictly positive). As already pointed out, such a scenario is an extremal case of superadditivity: two  objects containing no resource  of a given type, if put together constitute a single object that, surprisingly,  turns out to contain some amount of the resource. For this type of effect on the ground of quantum
channel capacities see \cite{SmithYard2010}.
 
 One can show \cite{DSSTT00} that the existence of an $n$-copy non-distillable state  is equivalent to the existence of a completely positive map $\Lambda$ such that it is  not completely co-positive but $2$-co-positive and its n-th tensor power,
 $\Lambda^{\otimes n}= \Lambda \otimes \cdots \otimes \Lambda$, is also $2$-co-positive. 
 Interestingly, there is also sufficient condition for existence of  NPT bound entanglement expressed in the lanquage of positive maps.
 If there exists a positive map 
$\Lambda$ that is neither completely positive, nor completely co-positive such 
 that its tensor power $\Lambda^{\otimes n}$ is positive for any $n$, 
 (this property is called {\it tensor-stable positivity}),
  then there exist  NPT bound entangled states
   that can be constructed explicitly on the basis of this map \cite{Muller-Hermes16}.

Note that for any $n$ there exists an $n$-copy non-distillable state that is $(n+1)$-copy distillable -- see \cite{Watrous04}. This fact  might be considered as an indication that the present problem of existence of NPT bound entanglement is hard.

  An important practical  observation being a footbridge between the 
 fourth and the fifth problem is contained in the following theorem \cite{HH99}: 
{\it NPT nondistillable entanglement exists if and only if there exist 
NPT nondistillable entangled Werner states for local dimension $d>2$.} 
In the smallest dimension, $d=2$, this is not the case.

This theorem implies that the current, fourth problem can be formulated as follows: 
{\it Decide, whether there exists an NPT Werner state which is nondistillable.} 

Werner states \cite{Werner1989} constitute a one-parameter family $\rho(d,{\alpha})$
of density matrices of order $d^2$  described in the next section. 
These states are NPT for $\alpha \in [-1,-\frac{1}{d})$ and 
 they are 1-copy nondistillable for  $\alpha \in [-\frac{2}{d},1]$.
 In this range of the parameter $\alpha$ 
they are conjectured to be nondistillable \cite{DCLB00,DSSTT00,Clarisse06PhD,Do16,Pankowski2010}.

Although the above theorem reduces the problem of  NTP bound entangled states 
to the question concerning a single-parameter family, it is not clear,
 whether analysis of this particular family of states provides the easiest technical way to solve the problem.
Some other subfamilies of the set of NPT states were also
 considered in the above context \cite{DSSTT00,Watrous04}. 

\subsection{ Distillability of quantum entnaglement}

{\bf Problem 5:} {\sl Show that the Werner state  $\rho(4, -1/2)$ of two ququarts,
$d=4$, defined in Eq. (\ref{Werner}) below, is not $2$-copy distillable.}
\bigskip

{\bf  Setup.} 
Consider the family of Werner states  defined on the Hilbert space $ {\cal H}_d \otimes  {\cal H}_d $
as  
\begin{equation}\label{Werner}
\rho(d,{\alpha})=\frac{\1 \otimes \1 + \alpha V}{d^{2} + \alpha d},
\end{equation}
 with 
the general range of the parameter  $\alpha \in [-1,1]$. The matrix $V$ stands for 
the {\it Swap operator}, defined by its matrix elements, 
 $\langle i j | V | k l \rangle= \delta_{i l} \delta_{j k}$. 
 Let us repeat that the above states are NPT for $ \alpha \in [-1, - \frac{1}{d}) $ and our fourth problem 
 can be just reduced to the analysis of their distillability in the cases of $d>2$.  
A distinguished state of this family,  $\rho(4, -1/2)$,  appearing in the 
problem considered here, is the only two-ququart Werner state 
such that its partial transpose (see Eq. \ref{NPT}) is proportional to a unitary matrix.

\begin{figure}[h]
	\centering
	\includegraphics[width=0.95\linewidth]{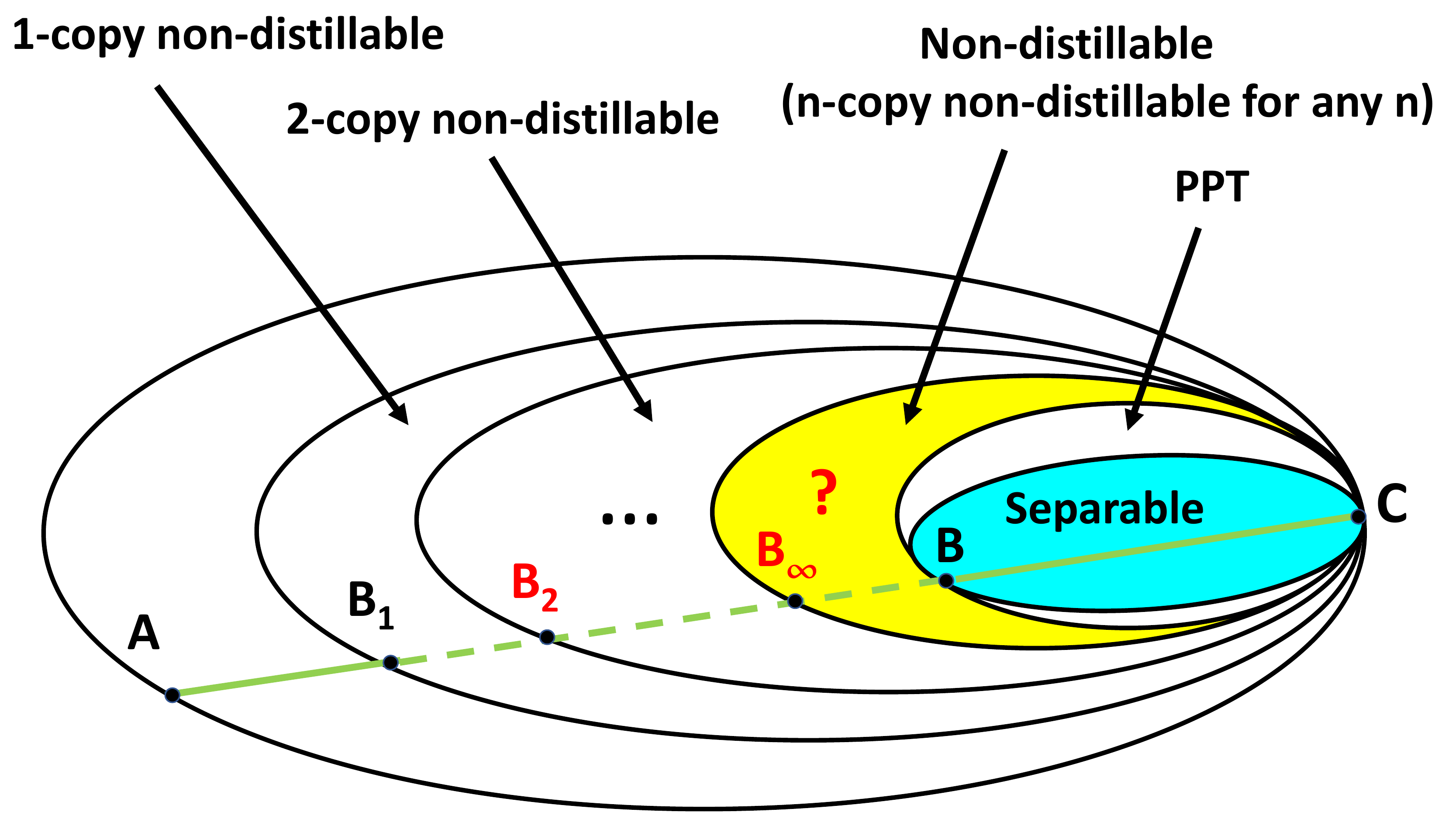}
	\caption{ A sketch of the convex set of mixed quantum states for a  $d \otimes d$ system
	  with $d>2$, 
	which contains the sets
	  of separable states, a larger set of PPT states and the sets of states with various classes of distillability.
	 The line represents the family of Werner states (\ref{Werner})
	 and the points $A$, $B_1$, $B$ and $C$  correspond to states labeled by  
	$\alpha$ equal to $-1$, $-\frac{2}{d}$, $-\frac{1}{d}$ and $1$, respectively. 
	 Point $A$ represents here the mixed state equal to 
	the normalised projector onto the antysymmetric subspace. 
     {\bf Problem~4} of existence of NPT bound entanglement is equivalent  
     to the question, whether  point $B_{\infty}$, the position of which is stil unknown,
      differs from point $B$.  
       The nature of the states along  the dashed-line $B_1 B$ is still unclear --
      to solve {\bf Problem~5} one has to decide,  
     whether  in case $d=4$ the unknown point $B_2$ is identical with $B_1$.}
	\label{fig:Werner-NPT-bound-entanglement}
\end{figure}

The Werner  states are 
  invariant with respect to twirling with local unitaries \cite{Werner1989},
  so they are also called $U \otimes U $-invariant.  
It was conjectured \cite{DCLB00,DSSTT00,Clarisse06PhD,Do16,Pankowski2010}
that the Werner states which are not $1$-copy distillable are just 
not-distillable, so in particular they are also $2$-copy non-distillable. 
The Problem  5 is visualized in Fig. \ref{fig:Werner-NPT-bound-entanglement},
and can be reduced to the question whether in the case $d=4$
the points $B_1$ and $B_2$ in this plot are equal.
 A stronger conjecture 
 that for Werner states 1-copy non-distillability is equivalent to complete non-distillability, means  that  $B_{\infty}=B_1$ in Fig. \ref{fig:Werner-NPT-bound-entanglement}.
\medskip

 {\bf Motivation.}               
On the physical side working on this problem might bring a step towards a
 proof of existence of NPT bound entanglement discussed in Problem 4. 
 Note that there is no promise for that due to the examples of Watrous \cite{Watrous04}, 
  in which the corresponding nondistillability property holds for $n$ copies, 
  but not for $n+1$ copies.
Yet some hope might come from the fact that Watrous states do not have  full rank --  as opposed to the above Werner states --  and this particular property seems important for their distillability.
Furthermore, possible  positive answer  would render the most natural 
 distillation protocols with two-copy interactions at their first stage. 
 On the other side, the negative solution of the problem would allow to construct practical entanglement distillation scheme for remarkably noisy states. 

On a mathematical side, the positive solution of the problem would provide a very elegant completely positive map that is not completely co-positive, 
but it is $2$-co-positive and the tensor product of its two copies also possesses this property.  
Interestingly,  this question is also equivalent to the following algebraic problem \cite{Pankowski2010}:  Show that the sum of squares of the two largest singular
values is bounded by $1/2$ for
 any Kronecker (tensor)  sum, $A \oplus B= A \otimes \1 + \1 \otimes B$,
 where $A$ and $B$ denote  traceless matrices of size $4$  satisfying 
 ${\rm Tr}(A^{\dagger}A) + {\rm Tr}(B^{\dagger}B) =1/4$. 
The bound equal to $1/2$ has been proven
 \cite{Pankowski2010}
 under the additional assumption that $A$ and $B$ are normal,
 so they commute with their Hermitian conjugates. Recently 
 further progress has been announced in a form of a theorem stating that  
 the bound still holds if one of the matrices is made completely arbitrary 
 \cite{Progress2019}.

Moreover, the explicit parameters $d=4$ and $\alpha=-1/2$ appearing in the problem are of special interest
since they correspond to the case of:
 \begin{itemize}
\item[(i)] The minimal dimension for which this very special Werner state is 1-copy non-distillable. 
Namely, the state $\rho(d, -2/d)$ has its partial transpose proportional  
to the dichotomic unitary operator
$U=I - 2 |\psi_{+}\rangle \langle \psi_{+}|$, where 
$|\psi_+\rangle$
denotes the maximally entangled state defined in (\ref{maxent}).
A  dichotomic unitary operator by definition has eigenvalues $\pm 1$.
\item[(ii)] The unique dimension, for which the above Werner state characterized by the parameter  $\alpha=-2/d$ is located  just on the boundary 
of a 1-copy non-distillability. For $d=4$  all the states with $\alpha < - 1/2$ are already 1-copy distillable,
which is not true for $d > 4$.
\end{itemize}

The choice of  the state with its partial transpose proportional 
to the  dichotomic unitary  operator
 is additionally motivated by the fact  that  checking its $n$-copy distillability seems to be easier
 than in the general case.
  In particular, the property of proportionality to the dichotomic unitary operation 
 is preserved with respect to taking the tensor product. 
\medskip

\subsection{ Further perspectives II}

 Solution of the fifth problem would likely provide us  some 
new insight into general properties of the Kronecker sum of  matrices,
$A \oplus B=A \otimes \1 + \1 \otimes B$, 
relevant from the point of view of quantum theory and
already studied in general matrix analysis \cite{Topics}. 
 Problem 4 seems to be rather complex, so its solution
 may involve some novel techniques concerning tensor products of several matrices.
Its positive solution would likely stimulate research on 
an important question relevant for quantum communication, namely, whether all 
NPT entangled states represent quantum privacy that may be 
distilled to the so-called private bit states  -- see \cite{HHHO05}.
If the answer were positive, the practical question would be,
whether such a distillation procedure can be achieved in the scheme
 called `one-way classical communication', 
in which one party, say Alice, communicates classical
bits to Bob, and not vice versa.

\section{Concluding remarks}

The goal of this article and the competition announced 
is to stimulate further reasearch on interesting mathematical 
problems directly related to quantum information applications.
Each problem described above has in a way been associated to a single
simple equation, 
which played a profound role in the development of the theory of quantum information.
Furthermore, each problem is illustated with a single figure, 
aimed to visualize the question posed. 

The problems concerning the discrete Hilbert space,
 related to deep algebraic and geometric
properties of  the set of quantum states,
are also linked to fundamental problems
from various branches of mathematics 
ranging from group theory  to number theory.
Solving some of them will impact a novel emerging field
of `quantum combinatorics'  --  the research on
existence and enumeration of various constellations of quantum states,
which satisfy certain conditions of balance and symmetry.

Results in the outlined directions, relevant to
our understanding of foundations of quantum theory,
can be useful for the development of quantum information processing.
Furthermore, some measurement schemes and
particular constellations of quantum states can
influence the computer designed quantum experiments \cite{KEZ20},
which may allow one to cope with a
huge number of  possible configurations which `explodes combinatorically'.
On the other hand, it is also thinkable that  the future results of new 
physical experiments designed in this way could bring hints concerning
some of the theoretical problems discussed in this work.

Possible solutions of problems devoted to distillability of quantum entanglement
 definitely would enrich our understanding of the nature of the tensor product -- 
 one of the key ingredients of quantum advantage in information processing. 
This knowledge might be relevant for development of  secure quantum communication 
in quantum networks. For instance NPT bound entanglement can 
lead to private bits that will allow to go beyond limitations known 
in quantum repeaters for PPT states with quantum security -- see \cite{Bauml15}.
Independently, the solution will shed a new light on important structures
known in linear matrix algebra.

\medskip

Finding a correct answer to any of the five problems presented above 
will be rewarded  by the Golden KCIK Award\footnote{The KCIK Award 
conferred for the year 2021 is set to 2021 EUR. 
If a given problem is not solved during
the year XXXX, the competition will automatically be extended to the next Year XXXX + 1 with the same rules and the  Award upgraded linearly to XXXX + 1 EUR.}
 established 
by the National Quantum Information Centre (KCIK) in
Poland. Each year up to two prizes can be awarded \cite{KCIK}.
The Competition will be closed if all five problems are solved. 
Then, as always happens, new problems will come into play...

\begin{acknowledgements}
We are pleased to thank Ingemar Bengtsson, Adam Burhardt, 
Dardo Goyeneche, Markus Grassl,   Otfried  G{\"u}hne,
Micha{\l} Horodecki, Felix Huber,  J\k{e}drzej Kaniewski, 
Arul Lakshimnarayan, Marcin Marciniak, Wojciech S{\l}omczy{\'n}ski
and Oliver Reardon-Smith for numerous helpful remarks and suggestions
and  to  Jakub Czartowski for preparing two figures.
 P.H. and {\L}.R. would like to acknowledge support by the
Foundation for Polish Science (IRAP project, ICTQT, Contract
No. 2018/MAB/5, cofinanced by the EU within the
Smart Growth Operational Programme),
while K.{\.Z} acknowledges support
by Narodowe Centrum Nauki under the grant number DEC-2015/18/A/ST2/00274 
and by the Foundation for Polish Science under the project 
Team-Net NTQC number 17C1/18-00.

\end{acknowledgements}


\begin{thebibliography}{200}
\bibitem{BP} Quantum version of bidding in bridge 
	 was analyzed in \cite{MTK14+}.
	  However, the cards used in this variant of the game
	   were entirely classical, in contrast to the `quantum cards' 
	    required for the problem discussed here.
\end{thebibliography}
\end{document}